\documentclass[prl,twocolumn,superscriptaddress,showpacs]{revtex4-2}

\usepackage{amsthm,amsmath,amsfonts,graphicx,bm}
\usepackage{bm}
\usepackage[dvipsnames]{xcolor}
\usepackage{bm,eufrak}
\usepackage[export]{adjustbox}
\usepackage{amsfonts}
\usepackage{amssymb}
\usepackage{amsmath,mathrsfs}
\usepackage[pdftex]{hyperref}
\usepackage{epsfig}
\usepackage{graphicx}
\usepackage{enumerate}
\usepackage{multirow}
\usepackage{verbatim}
\usepackage{subfigure}
\usepackage{bbold}
\usepackage{soul}
\usepackage{scrextend}
\usepackage{tikz}
\usepackage{comment}
\allowdisplaybreaks
\usepackage{hyperref}
\usepackage{float}

\newcommand{\ket}[1]{\vert#1\rangle}
\newcommand{\bra}[1]{\langle#1\vert}
\newcommand{\braket}[2]{\langle#1\vert#2 \rangle}

\newcommand{\tr}{\mathrm{tr}}

\newcommand{\mr}[1]{\mathrm{#1}}

\hypersetup{
  colorlinks   = true, 
  urlcolor     = Blue, 
  linkcolor    = BrickRed, 
  citecolor   = PineGreen 
}

\newcounter{notes}
\stepcounter{notes}

\DeclareFontFamily{OT1}{pzc}{}
\DeclareFontShape{OT1}{pzc}{m}{it}{<-> s * [1.10] pzcmi7t}{}
\DeclareMathAlphabet{\mathpzc}{OT1}{pzc}{m}{it}

\def\tcm{T.C.M. Group, Cavendish Laboratory, University of Cambridge, J.J. Thomson Avenue, Cambridge, CB3 0HE, UK}
\def\DAMTP{DAMTP, University of Cambridge, Wilberforce Road, Cambridge, CB3 0WA, UK}

\begin{document}

\title{Many-body-localization protection of eigenstate topological order in two dimensions}

\date{December 2022}

\author{Florian Venn}
\affiliation{\DAMTP}
\author{Thorsten B. Wahl}
\affiliation{\DAMTP}
\author{Benjamin B\'eri}
\affiliation{\DAMTP}
\affiliation{\tcm}

\begin{abstract}
Many-body localization (MBL) has been proposed to enable and protect topological order in all eigenstates, vastly expanding the traditional ground-state setting. 
However, for the most intriguing case of two-dimensional (2D) systems with anyons and topology-dependent degeneracies, the dense many-body spectrum challenges studying this MBL protection numerically.
Here we use large-scale full-spectrum variational ans\"atze to demonstrate MBL-protected topological order in the disordered 2D toric code perturbed by magnetic fields. 
We show that the system has topological local integrals of motion (tLIOMs) for magnetic field strengths below $h_c\approx0.1$ times the toric code coupling scale. 
Combining tLIOMs with exact diagonalization, we also identify high-energy topological multiplets in the dense many-body spectrum.
The phase diagram we find is consistent with toric-code and trivial MBL phases being separated by an intervening thermal phase.
\end{abstract}

\maketitle

\section{Introduction}
A key feature of many-body localization (MBL) is that eigenstates of MBL systems have low entanglement: they obey the area law~\cite{Fleishman1980,gornyi2005interacting,basko2006metal,znidaric2008many,pal2010mb,Bardarson2012,imbrie2016many,NandkishoreHuse_review,AltmanReview,Abanin2017,Alet2017,ImbrieLIOMreview2017,Friesdorf2015}. 
This is a commonality with topological ordered (TO) ground states~\cite{Kitaev_double,toric_code,WenPhysRevLett.90.016803,Kitaev2006,Levin_Wen,topQC,Kitaev_Preskill06,Levin_Wen2,Verstraete06}. 
MBL has indeed been proposed to expand TO to ``eigenstate TO", with TO appearing, and being protected, across the entire spectrum~\cite{Huse2013LPQO,2013Bauer_Nayak,topMBL,kjall2014many,bahri2015localization,2015Slagle,Thorsten,1DSPTMBL,2DSPTMBL,topMBL1D}.

In its most intriguing form, eigenstate TO generalizes 
two-dimensional (2D) intrinsic TO, i.e., TO with anyons and topological ground-state degeneracy~\cite{Kitaev2006,Levin_Wen,topQC,2013Bauer_Nayak,Huse2013LPQO,topMBL}. 
While much is known about the properties of such intrinsic
eigenstate TOs~\cite{2013Bauer_Nayak,Huse2013LPQO,topMBL}, their MBL protection, i.e., robustness against perturbations, is yet to be demonstrated. 
(As 2D MBL is expected to be only a metastable, albeit very long-lived, state~\cite{chandran2016higherD,deRoeck2017Stability,Potirniche2019,Gopalakrishnan2019,Doggen20}, we also mean robustness in this sense.)
This is a challenging problem. 
For example, the standard probe of ground-state TO via overlaps of exact and trial states and counting ground states on the torus does not generalize well: while the former is just hard to do across the spectrum, the latter is fundamentally challenged by high-energy topological multiplets (generalizing ground-state degeneracy) being split beyond the many-body level spacing which seemingly precludes their detection~\cite{Parameswaran2018}.

Here we use a large-scale full-spectrum variational approach to demonstrate eigenstate TO and its MBL protection in the 2D toric code---the simplest intrinsic 2D TO~\cite{Kitaev_double,toric_code,WenPhysRevLett.90.016803}. 
Our approach works by characterizing MBL via local integrals of motion (LIOMs)~\cite{chandran2015constructing,Rademaker2016LIOM,Abi2017,Goihl2018}: exponentially localized operators commuting with the Hamiltonian and each other. 
Specifically, we construct topological LIOM (tLIOM)~\cite{topMBL} and trivial LIOM ans\"atze from shallow quantum circuits~\cite{Wahl2017PRX} and topologically inequivalent sets of local ``stabilizers"---the skeletal (t)LIOMs describing the zero-correlation-length limit~\cite{Kitaev2006,Levin_Wen,topQC,topMBL}.  

Using this approach, supplemented by exact diagonalization (ED) to estimate phase boundaries, we show that the disordered toric code, perturbed by magnetic fields, is MBL with tLIOMs for magnetic field strengths below $h_c\approx0.1$ times the toric code coupling scale. 
This tLIOM evidence for the presence and MBL-protection of intrinsic eigenstate TO is from system sizes well beyond the reach of ED, thanks to MBL implying polynomial scaling for our (t)LIOM method~\cite{Wahl2017PRX,topMBL,topMBL1D}.
We also show that tLIOMs and ED can be synergestically combined: we use the two in tandem to detect high-energy topological multiplets in the system---even if these are buried by the dense many-body spectrum. 

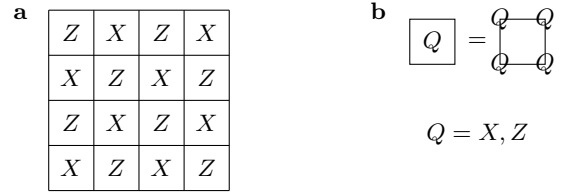
\begin{figure}[t]
	\begin{picture}(200,75)
		\put(0,0){
			\begin{tikzpicture}[scale=0.6]
				
				\draw (0,0)--(1,0)--(2,0)--(3,0)--(4,0);
				\draw (0,1)--(1,1)--(2,1)--(3,1)--(4,1);
				\draw (0,2)--(1,2)--(2,2)--(3,2)--(4,2);
				\draw (0,3)--(1,3)--(2,3)--(3,3)--(4,3);
				\draw (0,4)--(1,4)--(2,4)--(3,4)--(4,4);
				
				\draw (0,0)--(0,1)--(0,2)--(0,3)--(0,4);
				\draw (1,0)--(1,1)--(1,2)--(1,3)--(1,4);
				\draw (2,0)--(2,1)--(2,2)--(2,3)--(2,4);
				\draw (3,0)--(3,1)--(3,2)--(3,3)--(3,4);
				\draw (4,0)--(4,1)--(4,2)--(4,3)--(4,4);
								
				\node at (0.5,3.5){$Z$};
				\node at (2.5,3.5){$Z$};
				
				\node at (1.5,2.5){$Z$};
				\node at (3.5,2.5){$Z$};
				
				\node at (0.5,1.5){$Z$};
				\node at (2.5,1.5){$Z$};
				
				\node at (1.5,0.5){$Z$};
				\node at (3.5,0.5){$Z$};
				
				\node at (0.5,0.5){$X$};
				\node at (2.5,0.5){$X$};
				\node at (1.5,1.5){$X$};
				\node at (3.5,1.5){$X$};
				\node at (0.5,2.5){$X$};
				\node at (2.5,2.5){$X$};
				\node at (1.5,3.5){$X$};
				\node at (3.5,3.5){$X$};

				\begin{scope}[shift={(2,-0.2)}]
					\draw (6,3)--(7,3)--(7,4)--(6,4)--cycle;
					\node at (6.5,3.5){$Q$};
					\node at (7.5,3.5){$=$};
					\draw (8,3)--(9,3)--(9,4)--(8,4)--cycle;
					\node at (8,3){$Q$};
					\node at (9,3){$Q$};
					\node at (9,4){$Q$};
					\node at (8,4){$Q$};
				\end{scope}
				
				\begin{scope}[shift={(3,0)}]
					\node at (6.5,1.25){$Q=X,Z$};
				\end{scope}
				
			\end{tikzpicture}
		}
		\put(-10,65){\textbf{a}}
		\put(125,65){\textbf{b}}

	\end{picture}
	\caption{a: $4\times 4$ toric code with qubits on the vertices. Opposite sides are identified: the system forms a torus.  
		b:~The stabilizers $\prod_{i\in P_Q} Q_i$ for plaquettes $Q =X,Z$.
	} 
	\label{fig:toriccode}
\end{figure}

\section{Model}
We consider an $N_1 \times N_2$ square lattice of qubits (mostly taking $N_1=N_2\equiv N$) with Hamiltonian
\begin{align}
H = -\sum_{P\in P_X} J_P A_P -\sum_{P\in P_Z} J'_P B_P + \sum_i h_i Z_i. \label{eq:Ham}
\end{align}
The first two terms define the disordered toric code~\cite{toric_code}: we divide the set of plaquettes $P$ into $P_X$ and $P_Z$ plaquettes in a checkerboard pattern, and use $A_P = \prod_{i \in P} X_i$ and $B_P = \prod_{i \in P} Z_i$ (with $X_i$, $Y_i$, $Z_i$ Pauli matrices on site $i$), see Fig.~\ref{fig:toriccode}. 
The last term incorporates magnetic fields $h_i$. 
Unless otherwise stated, $J_P$, $J_P'$, and $h_i$ are random, drawn from normal distributions with zero mean and standard deviation $1$, $1$, and $\sigma$, respectively~\cite{Jpfn}. 
Our system, with qubits at sites instead of links, is $45^\circ$ rotated compared with the standard toric code to yield the simplest gate layout for our shallow circuits.
We use periodic boundary conditions (i.e., a torus); $N$ is even.

Taking $\sigma = 0$ leaves the disordered toric code. With $J_{P}^{(\prime)}$ disordered, this preserves MBL; 
this is the zero-correlation-length limit, with the mutually commuting ``stabilizers" $A_P$ and $B_P$ serving as tLIOMs. 
On a torus, the stabilizers are not all independent; a complete set of integrals of motion also includes the Wilson loops $\mathcal{Z}_k = \prod_{i \in \mathcal{C}_k} Z_i$ with  $\mathcal{C}_k$ ($k = 1,2$) along each of the torus' two inequivalent noncontractible loops~\cite{Kitaev_double,toric_code,WenPhysRevLett.90.016803}. 
Since the entire spectrum consists of toric-code stabilizer eigenstates, the system displays eigenstate TO~\cite{2013Bauer_Nayak,Huse2013LPQO,topMBL}. 
One of our key questions is how robust this eigenstate TO is for $\sigma \neq 0$: 
Upon increasing $\sigma$ we must eventually reach a topologically trivial phase, since upon $\sigma \rightarrow \infty$ we approach another zero-correlation-length limit with the $Z_i$ as stabilizer LIOMs, implying trivial product eigenstates.  

If there is a robust topological MBL phase, this must be separated from the trivial one either by a critical point or a thermal phase~\cite{1DSPTMBL,topMBL}. In 1D, theoretical studies suggest a thermal phase between topologically distinict MBL phases~\cite{Moudgalya20,Sahay21,topMBL1D,Laflorencie22,Jeyaretnam23}. 
Our other key question is whether such a phase appears in 2D: that would qualitatively distinguish eigenstate and ground-state phase diagrams, as the latter features a critical point separating trivial and topological states~\cite{Pich98,Motrunich2000,Kang2020}.

\section{Exact Diagonalization}  
We start with an ED analysis, leveraging that, by our judicious choice of the $h_i$ terms, $[B_P, H] = [\mathcal{Z}_k,H]= 0$: the $B_P$ thus remain tLIOMs (and the $\mathcal{Z}_k$ non-local integrals of motion) even for $\sigma\neq0$. 
The $B_P$ terms thus yield mere energy shifts; in our Ising considerations, we focus on $B_P\to 1$ and set $J'_P=0$, for concreteness. 
These Ising considerations arise from a duality of $H$, now describing $\mathbb{Z}_2$ gauge theory, to the 2D transverse field Ising model~\cite{Kogut1979,Polyakov2018}---a problem with only half the number of qubits.
In this duality, we replace $P_X$ plaquettes by qubits (with Pauli operators $X_P$, $Y_P$, $Z_P$) and map $A_P\to Z_P$ and $Z_i\to X_PX_{P'}$ (with $P$ and $P'$ the $P_X$ plaquettes adjacent to site $i$). 
These mappings preserve the algebra of operators.

Since the Ising model does not have TO, care is needed to preserve topological information: 
we map the four topological sectors, labeled by the $z_k=\pm 1$ eigenvalues of $\mathcal{Z}_k$, to four Ising models (see Appendix~\ref{app:duality}) by tracking the action of 
$\mathcal{X}_k = \prod_{i \in \mathcal{C}_{\bar k}} X_i$,
conjugate to $\mathcal{Z}_{k}$~\cite{Xjfn} 
[where ${\bar 1}=2$ and ${\bar 2} = 1$]
via $h_i\to h_{P P'}^{(z_1,z_2)}= 
h_i z_1^{\mathcal{C}_2(i)} z_2^{\mathcal{C}_1(i)}$ where $\mathcal{C}_k(i) = 1$ if $i \in \mathcal{C}_k$ and $0$ otherwise (we also relabeled $i\to PP'$ via the adjacent plaquettes). The dual Ising Hamiltonians are thus
\begin{align}
{\overline H}^{(z_1,z_2)} = - \sum_{P} J_P Z_P + \sum_{\substack{\langle P, P' \rangle}} h_{P P'}^{(z_1,z_2)} X_P X_{P'}, \label{eq:Ising}
\end{align}
where $\langle P, P' \rangle$ denotes nearest neighbors.

\begin{figure}
\includegraphics[width=\columnwidth]{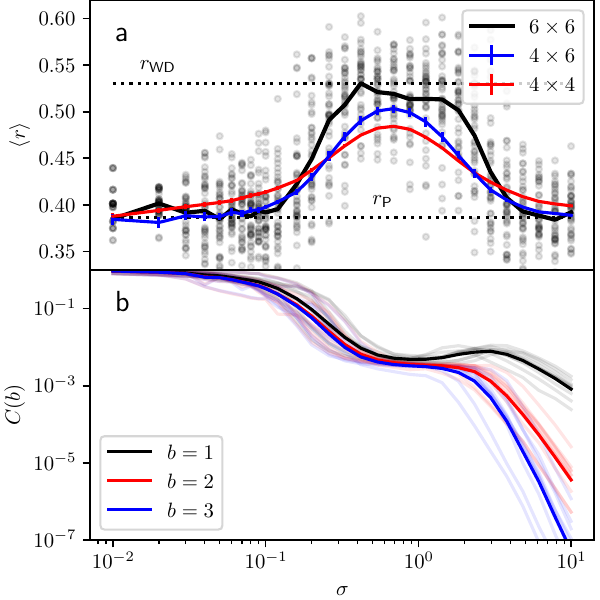}
\caption{ED results for $N=6$, 100 mid-spectrum states, and 10 disorder realizations, for a given (arbitrary) $B_P$ sector.
a: Gap ratio $r$, with disks for each disorder realization and topological sector (averaged over the 100 states) and the black line showing their average. 
Black dotted lines show $r_{\text{P}} = 0.386$ and $r_{\text{WD}} = 0.530$.
We also show the average $r$ for $N_1\times N_2=4\times 4$ (red, 2500 disorder realizations)
and $4\times6$ (blue, 200 disorder realizations) for the full spectrum. (Error bars:  $2\times$standard error.)
b: Wilson loop correlator $C(b)$ with faint lines for each disorder realization (geometrically averaged over the 100 states and four topological sectors) and thick lines showing their geometric  average.  
}
\label{fig:cor_gap}
\end{figure}

Our ED analysis is based on $\overline H$ for up to $N=6$ systems (cf. Fig.~\ref{fig:cor_gap}), for which we use 100 mid-spectrum eigenstates $\ket{\psi_n}$ of the Ising Hamiltonian~\eqref{eq:Ising} and their energies $E_n$, in each topological sector, for 10 disorder realizations. 
(We use mid-spectrum states to capture generic full-spectrum properties~\cite{Luitz2015,kjall2014many}.)
We first study the gap ratio $r = \min(\delta_n,\delta_{n+1})/\max(\delta_n,\delta_{n+1})$, where $\delta_n = |E_n - E_{n+1}|$, within each topological sector.
In an MBL phase, the lack of level repulsion implies $r_\mathrm{P} = 0.386$ (Poisson distribution); in a thermal phase level repulsion yields $r_\mathrm{WD} = 0.530$ (Wigner-Dyson distribution)~\cite{pal2010mb,Atas2013}. 
Fig.~\ref{fig:cor_gap}a shows the results: the gap ratio suggests two MBL phases, one for $\sigma\lesssim0.1$ and one for $\sigma\gtrsim5$. 
The data are consistent with a thermal phase between the two.
This is further corroborated by $N_1\times N_2=4\times 4$ and $4\times6$ systems: the $\sigma$ interval consistent with a thermal phase does not decrease with increasing system size.
Such a thermal phase in the topological MBL phase diagram is a key distinction from ground-state physics.
Our results suggest that such an intervening phase is a generic MBL feature, beyond the earlier 1D examples~\cite{Moudgalya20,Sahay21,topMBL1D,Laflorencie22}.

As the $\sigma\lesssim 0.1$ MBL phase includes $\sigma=0$, we expect it to be topological. 
Within ED, we test this via Wilson loops $\mathcal{W}_{\Omega}$, of area $|\Omega|$ and perimeter $|\partial\Omega|$, evaluated in $H$ eigenstates, $C_\Omega=|\langle n | \mathcal{W}_{\Omega} | n \rangle|$. 
For eigenstate TO this is expected to show a perimeter law $C_\Omega\propto e^{-c|\partial\Omega|}$, while for a trivial phase an area law $C_\Omega\propto e^{-c'|\Omega|}$  (here $c,c'>0$)~\cite{Wilson1974,Kogut1979,Polyakov2018,Parameswaran2018,2013Bauer_Nayak,Huse2013LPQO}. 
To test these asymptotics via the largest achievable loops, we study $\mathcal{X}_k(i) \mathcal{X}_k(i+b)$, with $\mathcal{X}_k(i)$ along Cartesian axis $k$,
at coordinate $i$ transverse to $k$. 
Hence, $|\Omega|=Nb$ and $|\partial\Omega|= 2N$ ($b\leq N/2$). 
We compute $C(b) = \frac{1}{2N}\sum_{ik}|\langle n | \mathcal{X}_k(i) \mathcal{X}_k(i+b) | n \rangle|$.
(For the Ising duality, we use $\mathcal{X}_k(i) \mathcal{X}_k(i+b)=\prod_{P\in \mathcal{P}_{i,b}}A_P$, with $\mathcal{P}_{i,b}$ the smaller set of plaquettes between the two loops, to replace $\ket{n}\to\ket{\psi_n}$ and  $\prod_{P\in \mathcal{P}_{i,b}}A_P\to \prod_{P\in \mathcal{P}_{i,b}} Z_P$.)
For fixed $N$, we expect a $b$-independent $C(b)$ for eigenstate TO, and $C(b)\propto e^{-c'Nb}$ in a trivial phase. 

Fig.~\ref{fig:cor_gap}b shows our results. 
$C(b)$ is  consistent with eigenstate TO for $\sigma\lesssim 0.1$ and a trivial MBL phase for $\sigma\gtrsim5$. 
However, the $b \leq 3$ range accessible via ED can substantiate a perimeter law only to a limited extent, preventing definite conclusions about the behavior of the Wilson loop correlator. This is why further innovation is needed to 
%
establish eigenstate TO for $\sigma\lesssim 0.1$. 
This could come from the topological quadruplets, whose splitting (mean absolute difference of energies in a quadruplet) should scale as $\Delta\propto \mathcal{O}(e^{-N/\xi_L})$, with $\xi_L$ the localization length~\cite{Huse2013LPQO,Parameswaran2018}.
However, away from low energies (e.g., near mid-spectrum), the typical level spacing is $\delta \propto \mathcal{O}(N^2 2^{-N^2})$ which can quickly obscure $\Delta$ in ED upon increasing $\sigma$ (and thus $\xi_L$) or $N$.

\section{Local Integrals of Motion} 
To overcome these ED limitations, we turn to our (t)LIOM ans\"atze. 
LIOMs are mutually commuting exponentially local operators $\tau_i$ ($i = 1, \ldots, N^d$) approximately fulfilling $[H, \tau_i] = 0$ and capturing $d$-dimensional MBL over its metastability lifetime $t_\text{th}$ and for system sizes $N<N_\text{th}$ where the entanglement area-law dominates in eigenstates~\cite{chandran2016higherD,deRoeck2017Stability,Potirniche2019,Gopalakrishnan2019,Doggen20}. 
(In terms of our parameters, $\ln t_\text{th}\sim N_\text{th} \sim \sigma^{-\mathrm{const.} \times (\ln \sigma)^2}\gg1$ for $\sigma \ll 1$~\cite{Gopalakrishnan2019}.)
LIOMs are usually dressed spins, e.g., $\tau_i = U S_i U^\dagger$, with $S_i=Z_i$ and $U$ a local unitary (mapping local to exponentially local operators).  
This implies an ansatz $\ket{\psi_{\bm s}}=U\ket{\bm{s}}$ for the MBL eigenstates ($\bm{s}=\{s_i\}$ are the $S_i$ eigenvalues). 
To describe TO MBL, this does not work because local unitaries cannot change TO~\cite{Bravyi2006,Chen_Gu,HastingsPRL2011} and, for $S_i=Z_i$, the $\ket{\bm{s}}$ are product states with trivial topology.
Instead, one must use tLIOMs with $S_i$ the local stabilizers for the TO (for the toric code, the $A_P$ and $B_P$)~\cite{topMBL}. 
On a topologically nontrivial manifold $\mathcal{M}$ (such as our torus), a complete set of integrals of motion also includes the nonlocal $\tau_k^\mathrm{nl} = U S_k^\mathrm{nl} U^\dagger$, with Pauli strings $S_k^\mathrm{nl}$  along noncontractible paths $\gamma_k$ in $\mathcal{M}$ [for Eq.~\eqref{eq:Ham}, $S_k^\mathrm{nl}=\mathcal{Z}_k$]. 
In an expansion of $H$ in terms of integrals of motion, the $\tau_k^\mathrm{nl}$ appear with coefficients exponentially small in the path lengths $|\gamma_k|$. 
Numerically, we can thus approximate $U$ by minimizing~\cite{Wahl2017PRX,2DMBL}
\begin{align}
f = \frac{1}{2} \sum_{i = 1}^{N^d} \tr([\tau_i^\mathrm{SR}, H][\tau_i^\mathrm{SR},H]^\dagger), \label{eq:fom}
\end{align}
where $\tau_i^\mathrm{SR} = U_\mr{SR} S_i U_\mr{SR}^\dagger$ with $U_\mr{SR}$ a shallow quantum circuit. 
By inspecting which stabilizer set minimizes $f$ (requiring also the minimal $f$ to be small for consistence with MBL~\cite{ellscfn}), one can distinguish different topological MBL phases.
In this way, since $\tau_i$ (and $\tau_k^\mathrm{nl}$) impact the entire spectrum, we get a full-spectrum probe of the MBL phase.
This approach proved successful at detecting topological MBL in 1D~\cite{topMBL1D}. Here, we explore its use in 2D to study MBL and TO for the toric code in Eq.~\eqref{eq:Ham}.

For $U_\mr{SR}$, we use a two-layer quantum circuit built of unitaries acting on $\ell \times \ell$ sites. 
To evaluate Eq.~\eqref{eq:fom}, similarly to Ref.~\onlinecite{2DMBL}, we use that each $\tau_i^\mathrm{SR}$ term yields a local tensor network contraction. 
For our 2D system, there are $N^2$ contractions; crucially the complexity is $N$-independent for each.
The unitaries making up $U_\mathrm{SR}$ can then be optimized by minimizing these local terms (each non-negative) using automatic differentiation~\cite{2DfMBL,1DSPTMBL}. 

We numerically optimized $U_\mathrm{SR}$ for $\ell = 2$, adapting Ref.~\onlinecite{2DMBL}'s algorithm for (t)LIOMs. 
We considered conventional LIOMs, with $S_i = Z_i$, and tLIOMs, with $S_i \in \{A_P, B_P\}$. 
Due to $[H, B_P] = 0$, we expect, and numerically found, that $[u,B_P] = 0$ for the $2\times2$-site unitaries $u$ (omitting position and layer indices) making up $U_\mathrm{SR}$~\cite{2DMBL,2DSPTMBL}. 
Hence, we must lay out the $u$ avoiding $[B_P,u]=0$ to imply $[Z_i,u]=0$  because that would preclude improving $f$ from its $U_\mathrm{SR} = \mathbb{1}$ value.
This is avoided by aligning the supports of $u$ and the $A_P$. 

\begin{figure}
\includegraphics[width=\columnwidth]{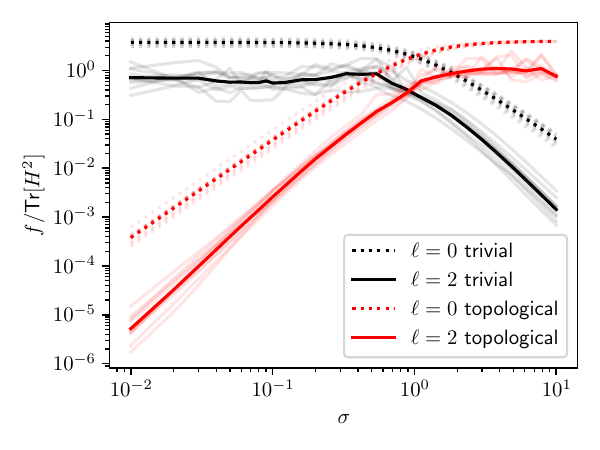}
\caption{Normalized $f$ [Eq.~\eqref{eq:fom}], both for trivial and topological LIOMs, for $\ell = 0$ and $\ell = 2$ (optimized), for an $N = 10$ ($100$-qubit) system. 
Faint lines show $f$ for 10 disorder realizations; thick lines show the disorder  average (geometrical).}
\label{fig:fom}
\end{figure}

We show the optimized $f$ in Fig.~\ref{fig:fom} for $\ell = 0, 2$ for both LIOMs and tLIOMs. 
($\ell = 0$ is $\tau_i^\mathrm{SR}=S_i$, i.e., $U_\mathrm{SR} = \mathbb{1}$.) 
As before, we use 10 disorder realizations, but now take $N=10$---well beyond the reach of ED. 
The results show that optimizing $U_\mathrm{SR}$ with $\ell = 2$ gives an order of magnitude improvement over $\ell = 0$. 
With $\ell = 2$, we find $f/\tr(H^2)<0.01$ for $\sigma\lesssim0.1$ and for $\sigma\gtrsim5$, consistent with the MBL phases seen in ED.  
Crucially, for $\sigma\lesssim0.1$, the tLIOMs yield two orders of magnitude smaller $f/\tr(H^2)$ than LIOMs (and vice versa for $\sigma\gtrsim5$). 
This provides clear large-scale evidence of eigenstate TO for $\sigma\lesssim0.1$ [and for its absence for $\sigma\gtrsim5$, but not for the precise phase transition point(s)~\cite{ellscfn}]:  
For $\sigma\lesssim0.1$ and $N\sim 10$, the scaling $N_\text{th} \sim \sigma^{-\mathrm{const.} \times (\ln \sigma)^2}\gg1$ makes rare regions unlikely~\cite{Gopalakrishnan2019} thus the system is MBL; since tLIOMs at fixed $\ell$ approximate this regime arbitrarily well as $\sigma$ is decreased (as indicated by an arbitrarily small $f$), while LIOMs do not, the corresponding (rare region-free) MBL phase must be TO.

\section{Topological quadruplets}
For $\sigma \rightarrow 0$, the Hamiltonian $H$ commutes with $\mathcal{X}_k, \mathcal{Z}_k$ ($k=1, 2$), resulting in topological quadruplets with degenerate energy. 
For nonzero $\sigma$, the $\mathcal{X}_k$ symmetry is lifted and the quadruplets are split.
Identifying which ED states belong to which quadruplet is easy if the splitting $\Delta$ is much smaller than the inter-quadruplet spacing $\delta$ (i.e., mean level spacing per topological sector): one can simply group states that are close in energy. 
However, this approach fails once $\Delta \gtrsim \delta$. 

We now show how tLIOMs may be used to identify quadruplets even after this point. 
We first note that the optimized tLIOMs $\tau_i^\mathrm{SR}$ imply the ansatz $\ket{\varphi_{\bm{s},\bm{z}}}=U_\mathrm{SR}\ket{\bm{s},\bm{z}}$, where $\ket{\bm{s},\bm{z}}$ are eigenstates of the $S_i\in\{A_P, B_P\}$ (with eigenvalues $\bm{s}=\{s_i\}$) and the $\mathcal{Z}_k$ (with eigenvalues $\bm{z}=\{z_k\}$). 
Our task is thus to identify which $\ket{\varphi_{\bm{s},\bm{z}}}$ approximates which ED eigenstate $\ket{n}$.

For this matching, we calculate $|\braket{n}{\varphi_{\bm{s},\bm{z}}}|^2$ and, given $\ket{n}$, select $\ket{\varphi_{\bm{s},\bm{z}}}$ that maximizes it. 
(In practice, we employ the Ising duality and map $\ket{\varphi_{\bm{s},\bm{z}}}\to\ket{\varphi^{\text{Ising}}_{\bm{s},\bm{z}}}$ in the Ising Hilbert space and compute  $\braket{\psi_n}{\varphi^{\text{Ising}}_{\bm{s},\bm{z}}}$ using the eigenstates $\ket{\psi_n}$ of $\smash{{\overline H}^{(\bm{z})}}$ in Eq.~\eqref{eq:Ising} with the corresponding $\bm{z}$. For a detailed explanation of the mapping see Appendix~\ref{app:duality}.)
Should multiple $\ket{n}$ match to the same $\ket{\varphi_{\bm{s},\bm{z}}}$, we consider the quadruplet labeled by $\bm{s}$ unidentified. 

\begin{figure}
\includegraphics[width=\columnwidth]{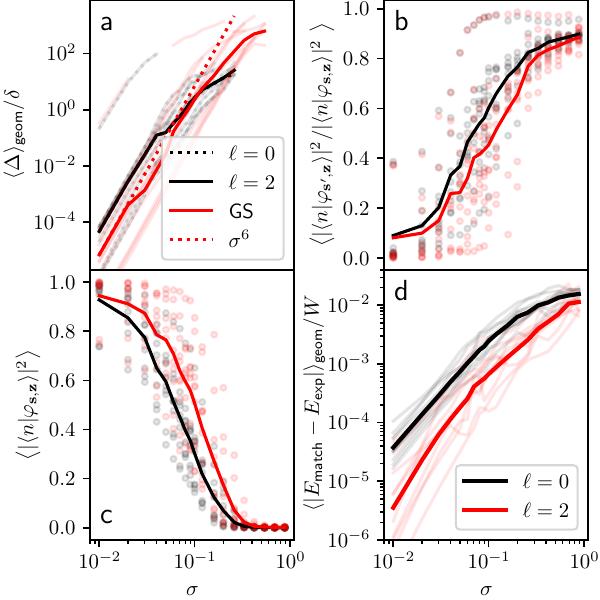}
\caption{a: Topological quadruplet splitting in the middle of the spectrum, from matching ED eigenstates $\ket{n}$ to tLIOM ans\"atze $\ket{\varphi_{\bm{s},\bm{z}}}$, using $\ell = 0,2$ circuits. 
There is negligible difference between the $\ell = 0$ and $\ell = 2$ results. 
We also show the ground-state quadruplet splitting (GS, for disorder realizations offering clear spectral separation) for comparison and a $\sigma^6$ line as a guide to the eye. 
In all panels, results from individual disorder realizations (10 realizations averaged over 100 states for mid-spectrum data) are shown faintly; solid lines show the disorder average. 
In panels with linear scale the averages are arithmetic, for logarithmic scales they are  geometric. 
All data are for $N=6$. 
In panels (b-d), $\ell=0$ is in black, $\ell=2$ is in red. 
b: The second best to best overlap ratio,  with $\ket{\varphi_{\bm{s}'\!,\bm{z}}}$ the second best and $\ket{\varphi_{\bm{s},\bm{z}}}$ the best match to $\ket{n}$. 
c: The overlap between the $\ket{n}$ and their matched $\ket{\varphi_{\bm{s},\bm{z}}}$. 
d: The difference between $E_{\text{exp}} = \bra{\varphi_{\bm{s},\bm{z}}} H \ket{\varphi_{\bm{s},\bm{z}}}$ and the exact ED energy, normalized to the spectrum width $W$.
}
\label{fig:splitting}
\end{figure}

Using this approach, we can track quadruplets in the middle of the spectrum into the $\Delta \gtrsim \delta$ regime; the resulting energy splittings are shown in Fig.~\ref{fig:splitting}a. 
(We again use $N=6$, and take 10 disorder realizations, and 100 mid-spectrum $\ket{n}$ from each topological sector.) 
For this matching task, using $\ell = 0$ turns out to be already sufficient: there is minimal difference between the obtained matches for $\ell=0,2$. 
To interpret why this is, we evaluate the ratio of the second largest to the largest $|\braket{n}{\varphi_{\bm{s},\bm{z}}}|^2$ for each $\ket{n}$ in the energy window considered. 
The results are shown in Fig.~\ref{fig:splitting}b.
We find that for $\sigma < 0.1$ this ratio is suppressed already for $\ell = 0$ (and more so for $\ell=2$); the matching task thus requires only a crude approximation to yield unique states to match to.

The splittings we obtain roughly follow the expected $\sigma^N$ scaling for small $\sigma$, similar to the ground-state splitting for the clean toric-code~\cite{Kitaev_double}. 
Deviations from this trend are expected because $\sigma^N$ is just a leading order (in $\sigma$) estimate. 
The deviations we observe are similar to those found for the ground-state (GS), also shown in Fig.~\ref{fig:splitting}a. 
The latter curve is a useful benchmark for our matching procedure because, unlike quadruplets mid-spectrum, it can typically be tracked directly due to the low-energy level spacing being sufficiently large to clearly separate this quadruplet from the rest of the spectrum.

An additional quantity the tLIOMs give access to is the expected energy $E_{\text{exp}} = \bra{\varphi_{\bm{s},\bm{z}}} H \ket{\varphi_{\bm{s},\bm{z}}}$. 
For $\ket{\varphi_{\bm{s},\bm{z}}}$ maximizing $|\braket{n}{\varphi_{\bm{s},\bm{z}}}|^2$, this should be close to the energy $E_{\text{match}}$ of the ED state $\ket{n}$, provided $|\braket{n}{\varphi_{\bm{s},\bm{z}}}|^2$ is large. 
We plotted this overlap in Fig.~\ref{fig:splitting}c and the difference between $E_{\text{exp}}$ and $E_{\text{match}}$ normalized to the spectrum width %
in Fig.~\ref{fig:splitting}d. 
We find that going from $\ell = 0$ to $\ell = 2$ improves the overlap, which yields roughly an order of magnitude improvement in the energy approximation, a similar effect was observed in the surface code as presented in Appendix~\ref{app:surface}.

\section{Conclusion}
We have demonstrated the presence and MBL-protection of eigenstate TO in the disordered perturbed toric code. 
Using tLIOMs constructed via stabilizers and shallow quantum circuits, we provided full-spectrum variational evidence for $N=10$, i.e., $100$-qubit systems. 
This is well beyond the reach of ED, even for the uniaxial magnetic fields we considered that allow one to exploit an Ising duality. 
Using that duality, however, we could combine tLIOMs with ED for $N=6$ (i.e., $36$-qubit 
systems) and detect topological multiplets in the dense many-body spectrum. 
Such multiplets have heretofore been thought to be undetectable.

The tLIOM and ED evidence, including the gap ratio in the latter case, revealed a phase diagram with toric-code and trivial MBL phases for $\sigma\lesssim0.1$ and $\sigma\gtrsim 5$, respectively. 
For $0.1\lesssim \sigma \lesssim 5$, our results are consistent with an intervening thermal phase. 
While such intervening thermal phases have been noticed between topologically distinct 1D MBL phases~\cite{Moudgalya20,Sahay21,topMBL1D,Laflorencie22}, our results suggest that they are more generic, appearing also in higher-dimensional topological MBL phase diagrams. 

Further studies could investigate eigenstate TO and the phase diagram for a broader class of systems, including the 2D toric code with more generic perturbations. 
The tLIOM approach is equally suitable for this: it requires only the TO to be nonchiral and Abelian so that a suitable stabilizer set exists~\cite{topMBL}. 
For these more general systems, we expect that instead of ED, our method can be combined with 2D extensions of the excited-state density matrix renormalization group~\cite{Khemani2016MPS,Pekker2017MPO,Devakul2017,Yu2017,2D_DMRG-X}, based, e.g., on efficiently contractible isometric tensor networks~\cite{IsoTN1,IsoTN2}. 

Finally, the robustness of MBL with TO we demonstrated may motivate progress towards realizing intrinsic eigenstate TO in experiments. 
Recent results on creating surface-code ground states in many-qubit systems~\cite{satzinger2021realizing,krinner2022realizing,Acharya22}, with control over the requisite $A_P$ and $B_P$, suggest that creating the MBL surface code (a planar version of the toric code we studied), or driven systems building on this phase, e.g., surface-code topological time crystals~\cite{TopTC}, may soon be within experimental reach.
Wilson loop expectation values, including the correlator $C(b)$ (Fig.~\ref{fig:cor_gap}b), are naturally accessible in such experiments, and, with access to a larger range of $b$, would provide compelling fingerprints for the MBL TO phase.

This work was supported by EPSRC grant EP/V062654/1, the Royal Society Research Fellows
Enhanced Research Expenses 2021 RF\textbackslash{}ERE\textbackslash{}210299, and in part by the ERC \mbox{Starting Grant No. 678795 TopInSy.}
Our simulations used resources at the Cambridge Service for Data Driven Discovery operated by the University of Cambridge Research Computing Service (\href{www.csd3.cam.ac.uk}{www.csd3.cam.ac.uk}), provided by Dell EMC and Intel using EPSRC Tier-2 funding via grant EP/T022159/1, and STFC DiRAC funding  (\href{www.dirac.ac.uk}{www.dirac.ac.uk}).

\appendix
\section{Duality Relation}
\label{app:duality}
Here, we review the duality between 2D $\mathbb{Z}_2$ gauge theory and the 2D transverse field Ising model~\cite{Kogut1979,Polyakov2018}, as applied to our Hamiltonian Eq.~(1). We also discuss how topological sectors are incorporated  by considering multiple Ising models, and how the tLIOM ans\"atze are mapped under the Ising duality. 
For clarity, here we denote Pauli operators of the toric code by $\sigma_i^\alpha$ ($\alpha=x,y,z$) and those in the Ising system by $X_P,Y_P,Z_P$.

We consider the Hamiltonian $H$ in an eigenspace of all $\{B_P\}$, where these have eigenvalues $\{b_P\}$. 
The eigenstates of all $A_P$'s and $\mathcal{Z}_{1,2}$ can be chosen as basis vectors of the considered eigenspace, $|a_1, a_2, \ldots, a_{N^2/2}; z_1, z_2\rangle$, $a_P = \pm 1, z_k = \pm 1$, with 
\begin{align}
	\!\!\!A_P |a_1, \ldots, a_{N^2/2}; z_1, z_2\rangle &= a_P |a_1, \ldots, a_{N^2/2}; z_1, z_2\rangle, \label{eq:A_P} \\ 
	\!\!\!\mathcal{Z}_k |a_1, \ldots, a_{N^2/2}; z_1, z_2\rangle &= z_k|a_1, \ldots, a_{N^2/2}; z_1, z_2\rangle.
\end{align}
We note that the relevant subspace of the space spanned by those basis vectors is given by $\prod_{P \in P_x} a_P = 1$, because $\prod_{P \in P_x} A_P = \mathbb{1}$ (similarly, $\prod_{P \in P_z} b_P = 1$). 

\begin{figure}
	\includegraphics[width=0.35\columnwidth]{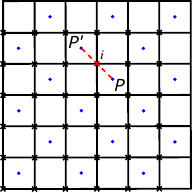}
	\caption{Mapping the toric code with $\sigma_i^z$ magnetic field terms (resulting in $\mathbb{Z}_2$ gauge theory) to the transverse field Ising model, exemplified for $N=6$. Crosses mark original sites and blue dots mark $P_x$ plaquettes, the latter becoming the sites for the Ising Hamiltonian $\overline H$. 
		Nearest-neighbor $P_x$  plaquettes uniquely define a site $i$ on the original lattice (red cross).}
	\label{fig:Ising}
\end{figure}

We now analyze how the individual terms in the Hamiltonian act on these basis states.
From Eq.~\eqref{eq:A_P} it follows that the $A_P$ term in Eq.~(1) acts as a Pauli-$z$ operator in the above basis, $-\sum_{P \in P_x} J_P Z_P$ (with Pauli-$z$ eigenvalues $\{a_P\}$). 
The second term of the Hamiltonian is diagonal, because we are in an eigenspace of all $\{B_P\}$; this term thus becomes $-\sum_{P \in P_z} J_P b_P = f(\{b_P\})$. 
Finally, $\sigma_i^z$ acts nontrivially on the two adjacent $P_x$ plaquettes, say $P$ and $P'$. 
Since these anticommute with $\sigma_i^z$, i.e., $\{A_P, \sigma_i^z\} = \{A_{P'}, \sigma_i^z\} = 0$, the last term in Eq.~(1) acts like $\sum_{i} h_{PP'} X_P X_{P'}$ in the new basis (see Fig.~\ref{fig:Ising}) where we relabeled $h_i\to h_{PP'}$ via the adjacent plaquettes. 

This yields a 2D transverse field Ising Hamiltonian $\overline H$, plus $f(\{b_P\})$. 
This $\overline H$ has a $\mathbb{Z}_2$ symmetry, $[\overline H, \prod_{P \in P_x} Z_P] = 0$, absent in the original Hamiltonian $H$.
Due to $\prod_{P \in P_x} a_P = 1$, the spectrum of $H$ [Eq.~(1)] corresponds only to the even symmetry sector of $\overline H$.

Conversely, $H$ has the symmetries $[H, \mathcal{Z}_k] = 0$ ($k = 1,2$), which do not exist in $\overline H$: since the Ising mapping takes $\sigma_i^z$ to adjacent $X_P X_{P'}$, it maps $\mathcal{Z}_k=\prod_{i\in \mathcal{C}_k}\sigma_i^z\to \mathbb{1}$. 
For an operator $\mathcal{O}$, such as $H$, depending on (products of) the $A_P$ and the $\sigma_i^z$, yielding  $[\mathcal{O},\mathcal{Z}_k]=0$, this implies the following: 
While $\mathcal{O}$ is block-diagonal in $z_1,z_2$, i.e., has four blocks (topological sectors) $\bra{z_1,z_2}\mathcal{O}\ket{z_1,z_2}$, the duality yields an Ising-Hilbert-space operator $\overline{\mathcal{O}}$ describing only the block $\bra{1,1}\mathcal{O}\ket{1,1}$ because this is the only block consistent with $\mathcal{Z}_1\to\mathbb{1}$. 
To recover the other blocks, we can, however, use that $\mathcal{X}_k=\prod_{i\in \mathcal{C}_{\bar k}}\sigma_i^x$ [where ${\bar 1}=2$ and ${\bar 2} = 1$]
is conjugate to $\mathcal{Z}_{k}$ and hence it toggles between topological sectors.
Thus, $\bra{z_1,z_2}\mathcal{O}\ket{z_1,z_2}=\bra{1,1}\mathcal{X}^{n_1}_1\mathcal{X}^{n_2}_2\mathcal{O}\mathcal{X}^{n_2}_2\mathcal{X}^{n_1}_1\ket{1,1}$, with $n_k=(1-z_{k})/2$. 
We can therefore reintroduce the other blocks by applying the Ising duality to $\mathcal{X}^{n_1}_1\mathcal{X}^{n_2}_2\mathcal{O}\mathcal{X}^{n_2}_2\mathcal{X}^{n_1}_1$ (another operator depending only on $A_P$ and the $\sigma_i^z$) to get four operators $\overline{\mathcal{O}}^{(z_1,z_2)}$.
This is the approach we use for $\mathcal{O}=H$. 
The four topological sectors thus yield four Ising Hamiltonians ${\overline H}^{(z_1,z_2)}$ with couplings modified according to 
$h_{P P'}\to h_{P P'}^{(z_1,z_2)}= 
h_i z_1^{\mathcal{C}_2(i)} z_2^{\mathcal{C}_1(i)}$ where $\mathcal{C}_k(i) = 1$ if $i \in \mathcal{C}_k$ and $0$ otherwise.
This gives the Ising Hamiltonians in Eq.~(2) of the main text, plus $f(\{b_P\})$.

In our ED computations we use that ${\overline H}$ (suppressing superscripts for brevity) being defined on a bipartite lattice leads to a spectral symmetry: 
denoting the set of sites belonging to each sublattice by $G$ and $G'$, the operator $R=\prod_{P\in G,P'\in G'} X_P Y_{P'}$ satisfies $R {\overline H} R^\dagger = -{\overline H}$. 
Therefore, the spectrum of each of the $B_P$-sectors is symmetric with respect to $f(\{b_p\})$, which for simplicity we set to zero in the main text. 
To reduce the dimension of the ED problem further we use that $[\overline H^2, R] = 0$ and solve only for eigenstates of $\overline H^2$ in the $R = 1$ sector. 
To get representative mid-spectrum states we solve for the 100 states with the largest eigenvalues of $\overline H^{-2}$, from which we can obtain, using the $R$ anti-symmetry, the eigenstates $\ket{\psi_n}$ and energies $E_n$ of 200 mid-spectrum states of $\overline H$. 
(Also by the $R$ anti-symmetry, either the 100 positive or 100 negative energy states can be used as independent entities for statistical purposes.)

In matching exact eigenstates $\ket{n}$ of $H$ to tLIOM ans\"atze $\ket{\varphi_{\bm{s},\bm{z}}}$, we must compute $\braket{n}{\varphi_{\bm{s},\bm{z}}}$, which we also do via Ising duality. 
(As in the main text, we use the notation ${\bm s}=\{s_i\}=\{a_i,b_i\}$ and ${\bm z}=(z_1,z_2)$ and take $b_i=1$ for concreteness.)
We replace ED eigenstates by Ising eigenstates $\ket{\psi^{(\bm z)}_n}$
of $\smash{{\overline H}^{(\bm{z})}}$, where we made the topological sector $\bm z$ explicit. 
To use these for overlaps, we must also map the  $\ket{\varphi_{\bm{s},\bm{z}}}$ to the Ising Hilbert space. 
For this, we use that due to $[U_\mr{SR},B_P]=0$, the expansion of $U_\mr{SR}$ in terms of Pauli operators can include only products of $\mathbb{1}$, and the various $A_P$ and $\sigma_i^z$ (i.e., $\sigma_i^x$ Pauli strings along any open paths are  forbidden because these would anticommute with the $B_P$ at their path endpoints). 
By requiring $U_\mr{SR}$ to preserve $\bm z$, i.e., $[U_\mr{SR},\mathcal{Z}_k]$, the operator $\mathcal{X}_k$ is also not allowed. 
This means that one can apply the duality mappings $A_P\to Z_P$ and $\sigma_i^z\to X_PX_{P'}$ (and $B_P\to b_P$) to these circuits. 
(In practice, we parameterize the gates of $U_\mr{SR}$ using matrix exponentials and apply the duality mapping in the exponents.)
As we were for  $\overline H$, we must be careful not to lose the topological sectors in this process. 
To this end, similarly to $\overline H$, we  apply the duality mapping to $\mathcal{X}^{n_1}_1\mathcal{X}^{n_2}_2 U_\mr{SR} \mathcal{X}^{n_2}_2\mathcal{X}^{n_1}_1$, which yields four distinct unitaries $\overline{U}^{(\bm z)}_\mr{SR}$ in the Ising Hilbert space.
The corresponding Ising-mapped tLIOM ans\"atze are
$\ket{\varphi^{\text{Ising}}_{\bm{s},\bm{z}}}=\overline{U}^{(\bm z)}_\mr{SR}\ket{\bm s}$, where $\ket{\bm s}$ are $Z_P$ eigenstates in the Ising Hilbert space. 
The overlaps we compute are hence $\braket{n}{\varphi_{\bm{s},\bm{z}}}\to \braket{\psi^{(\bm z)}_n}{\varphi^{\text{Ising}}_{\bm{s},\bm{z}}}$.

\section{Topological MBL in the surface code}
\label{app:surface}

\begin{figure}
\includegraphics[width=\columnwidth]{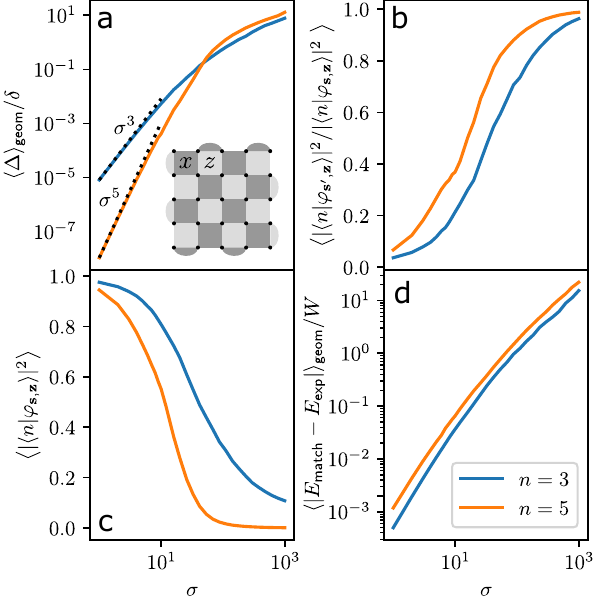}
\caption{Simulation of a square $n$-by-$n$ qubit surface code of size $n=3$ and $5$ with open boundaries in analogy to Fig.~4. a: Topological splitting from matching ED eigenstates $\ket{n}$ to $\ell = 0$ tLIOM ans\"{a}tze $\ket{\varphi_{\bf{s}, z}}$. Dashed lines indicate the expected $\sigma^n$ scaling. In all panels we have averaged over 100 disorder realizations, we use linear averages in plots with linear scales and logarithmic averages in plots with logarithmic scales. Inset: The configuration of the surface code we are using, here the 5-by-5 setting. We indicate the qubits as black dots and the $X$ ($Z$) stabilizers as dark (light) gray faces. 
b: The second best to best overlap ratio,  with $\ket{\varphi_{\bm{s}'\!,z}}$ the second best and $\ket{\varphi_{\bm{s},z}}$ the best match to $\ket{n}$. 
c: The overlap between the $\ket{n}$ and their matched $\ket{\varphi_{\bm{s},z}}$. 
d: The difference between $E_{\text{exp}} = \bra{\varphi_{\bm{s},z}} H \ket{\varphi_{\bm{s},z}}$ and the exact ED energy, normalized to the spectrum width $W$. \label{fig:surface}}
\end{figure}

Here, we study topological MBL in the disordered surface code with open boundary conditions. We use an $n$-by-$n$ qubit surface code in a configuration featuring weight-two stabilizers at the boundaries as shown in the inset in Fig.~\ref{fig:surface}a.
We analyse the same Hamiltonian as in Eq.~(1), but with the stabilizer plaquettes taken from the open boundary surface code.

For the analysis we again employ the duality between the surface code and the Ising model similar to the process described in Appendix~A. The only difference arises due to the fact that the surface code with open boundary conditions only features a single logical $\mathcal{Z}$ operator. Therefore, the resulting logical space is only two-dimensional and we obtain just two Ising Hamiltonians $\overline{H}^{(z)}$, that each describe one logical subspace. Since we focus on the 3-by-3 and 5-by-5 codes, the Hilbert space is significantly smaller and we do not require additional optimizations and can use ED to solve for the eigenstates of $\overline{H}^{(z)}$. The eigenstates of the Ising Hamiltonian also give us access to the eigenstates $\ket{n}$ of the disordered surface code.

We want to perform a matching of the ED states $\ket{n}$ to the topological ansatz functions. Since the logical space is two-dimensional, we now parameterize the ansatz functions as $\ket{\varphi_{{\bf{s}}, z}}$, where ${\bf{s}} = \{s_i\} = \{a_i, b_i\}$ are the eigenvalues of the stabilizers (again we only focus on $b_i = 1$) and $z$ is the eigenvalue of the single logical $\mathcal{Z}$ operator. We focus here only on $\ell = 0$, because the optimization of the unitary circuits has not brought significant improvement for the matching of the toric code (cf. Fig.~4a).

Using the ED states and ans\"{a}tze, we compute the overlap $|\braket{n | \varphi_{{\bf{s}}, z}}|^2$ and perform the same matching procedure as described in the main text. Employing the matching, we can determine the topological doublet splitting as shown in Fig~\ref{fig:surface}a. We find that the topological splitting of the $n$-by-$n$ surface code follows a $\sigma^n$ scaling, as expected. In analogy to Fig.~4 we also plot the average ratio between best and second best overlap (Fig.~\ref{fig:surface}b), the average overlap (Fig.~\ref{fig:surface}c), and the average difference between energy eigenvalue and expectation value of the ansatz (Fig.~\ref{fig:surface}d).


\begin{thebibliography}{75}
\bibitem [{Jpf()}]{Jpfn}%
  \BibitemOpen
  \href@noop {} {}\bibinfo {note} {The constants $J'_P$ will eventually be set
  to zero; we retain them here to present all the stabilizers, described
  below.}\BibitemShut {Stop}%
\bibitem [{Xjf()}]{Xjfn}%
  \BibitemOpen
  \href@noop {} {}\bibinfo {note} {The $\mathcal{X}_k$, unlike the
  $\mathcal{Z}_k$, are not integrals of motion.}\BibitemShut {Stop}%
\bibitem [{ell()}]{ellscfn}%
  \BibitemOpen
  \href@noop {} {}\bibinfo {note} {Studying MBL phase boundaries is possible,
  e.g., via the scaling of $f/\text{Tr}(H^2)$ with gate-length or circuit
  depth; we expect exponential decay for MBL and non-exponential decay for a
  delocalized phase. However, studying this requires circuits beyond our
  current reach.
  }\BibitemShut {Stop}%
\end{thebibliography}
\end{document}